\documentclass[reprint]{revtex4-1}
\pdfoutput=1
\usepackage{fullpage}
\usepackage{draftcopy}
\usepackage{graphicx}
\usepackage{verbatim}
\usepackage{color}
\usepackage{ulem}

\begin{document}

\title{Renyi Entanglement Entropy of Molecules: Interaction Effects and Signatures of Bonding}
\author{Norm M. Tubman}
\author{Jeremy McMinis}
\affiliation{Department of Physics, University of Illinois Urbana Champaign, Urbana, IL 61801}
\date{\today}
\begin{abstract}
Recent proposals of measuring bipartite Renyi entropy experimentally 
involve techniques that hold exactly for non-interacting quantum particles. Here we consider the difference between such measurements and the actual Renyi entropy for ground state fermion \textit{ab initio} molecular systems.  To calculate various entanglement measures we extended several different techniques for use with variational Monte Carlo, Hartree-Fock and quantum chemistry methods.  Our results show that in systems with strong electron correlations the Renyi entropy may not be accurately determined with the proposed measurements.  In addition, we find significant physical insight can be gained by calculating entanglement properties in molecular systems. We see that the Renyi entropy and the entanglement Hamiltonian encode information about the character of the covalent bonds in a molecular system and that such information may lead to better descriptions of bonded systems that have been traditionally hard to describe with standard techniques.  In particular our results for the C$_{2}$ dimer suggests all eight valence electrons play a significant role in covalent bonding.   
\end{abstract}

\maketitle

The study of bipartite entanglement entropy and other entanglement properties has led to new insights in electronic properties of extended systems as seen in many recent studies\cite{1dint-1,1dint-2,1dint-3,widom-1,widom-2,widom-3,nonint-1,nonint-2,dftent-1,op-1,review-1}.  
  Real space bipartite entanglement properties are derived from reduced density matrices in which space is divided into two regions and spatial degrees of freedom are traced out over one of these regions.  This suggests they contain information of how quantum systems are spatially connected to each other.  Interpreting entanglement properties to develop an understanding of how they are useful in describing quantum systems is still an active field.  This is in part due to the relatively small number of studies of entanglement properties in interacting systems that are not related to fractional quantum Hall studies or 1D Hamiltonians\cite{renyi-1,renyi-2,renyi-3,renyi-4}. 
  Understanding entanglement properties of interacting systems is of prime interest in recent proposals that suggest direct measurements of entanglement and Renyi entropies are experimentally possible\cite{measure-1,measure-2,measure-3,measure-4,measure-5}.

\begin{figure}[htp]
\centering
\fbox{\includegraphics[scale=0.3]{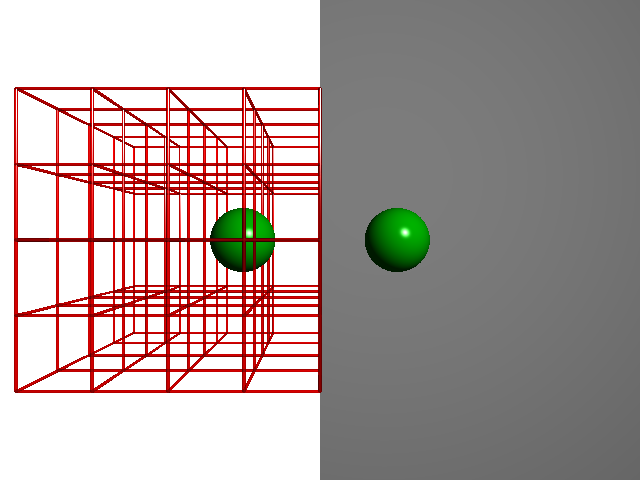}}
\caption{The bipartite entanglement properties are calculated on space partitioned into two half spaces for homonuclear diatomic molecules.  Region A can be considered the left half space, and region B, which is traced out, is the shaded region on the right.  The correlation method uses a large grid in region A, while the swap and the fluctuation methods are sampled continuously in region A to compute the Renyi entropies.     \label{fig:front}}
\end{figure}

We consider two specific experimental proposals for measuring entanglement properties in this work:  the correlation function approach\cite{redmat-1,corr-1,corr-2} and particle number fluctuations\cite{measure-1,measure-2}.  It has been shown that both of these methods are exact when the ground state wave function of interest represents a non-interacting system.  However it is unknown what biases these proposed measurements will have in determining entanglement for interacting systems.    
On a more fundamental level we consider the relationship between entanglement properties and the characterization of molecular systems.  In particular we explore the idea that the reduced density matrix, generated by tracing out real space degrees of freedom, contains information about the bonding properties of a system.  
In this work we develop the preliminary framework for using these tools to analyze covalent bonds. 

 The Renyi entropies are among several quantities of interest that are related to the reduced density matrix of a subsystem.  The reduced density matrix that describes a subsystem will in general have finite Renyi entropies which are given by
\begin{equation}
S_{n}(A) = \frac{1}{1-n}\ln[\textrm{Tr}((\rho_{A})^{n})]
\end{equation}
where $n > 0$ and $\rho_{A}$ refers to the reduced density matrix calculated when the degrees of freedom outside of region A are traced out from the density matrix $\rho$ of the entire system.  
 The Renyi entropies for integer values $n > 1$ are directly accessible from quantum Monte Carlo (QMC).
In this work we are primarily interested in S$_{2}$ and all further references to Renyi entropy will refer to S$_{2}$ except when explicitly stated otherwise.

We use three techniques for calculating the Renyi entropy, which we refer to as the swap method, correlation method and fluctuation method.  The swap method is an unbiased estimator, and is the benchmark for which the other two methods should be compared.  The other two methods involve calculating intermediate values that are closely related to quantities that in principle can be measured experimentally.  Through a combination of variational Monte Carlo (VMC) and quantum chemistry methods we can compute the Renyi entropy  of the fluctuation and correlation methods and compare to the value that can be computed with the swap method\cite{renyi-1}.  The quantities to be calculated in the particle fluctuation method are the moments, $\mu_{m}$, of the particle number operator in region A, $\mu_{m} = \langle \Psi | (\hat{N}_{A})^{m}) | \Psi \rangle /\langle \Psi |\Psi \rangle$.  With a finite number of moments it is possible to estimate S$_{n}$\cite{measure-1}.  Calculating these moments can be performed directly in VMC.  

The correlation method is performed in a quite different manner and involves the expectation value of specific single body operators. The expectation value of these operators can be calculated with the one particle reduced density matrix $\gamma_{1}$ using Hartree-Fock and configuration interaction techniques.  An important quantity that is calculated as an intermediate step in the correlation method is the entanglement Hamiltonian, $\mathcal{H}_{ent}$, which is defined as $\rho_{A} = \kappa e^{-\mathcal{H}_{ent}}$.  The normalization $\kappa$ is set to ensure Tr$(\rho_{A})=1$.  All three methods are exact for Hartree-Fock wave functions.  The Hartree-Fock results are shown in figure \ref{fig:c-ent} for the C$_{2}$ dimer. The details of these methods are described in the supplemental material.

\begin{figure}[htp]
\centering
\includegraphics[scale=0.4]{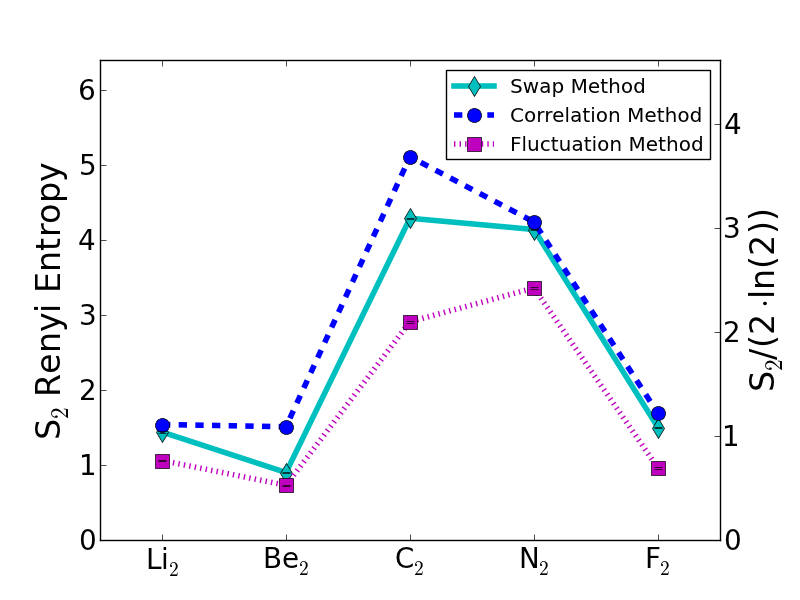}
\caption{The bipartite Renyi entropy, S$_2$, calculated for several diatomic molecules at their equilibrium geometries with the three different techniques described in the text.  The swap operator is an unbiased estimate while the fluctuation and correlation methods are what would be measured with experiment for the same ground state wave functions.    \label{fig:molecules}}
\end{figure}

The systems considered in this work are those of the homonuclear first row dimers with ground state wave functions that are known to be spin singlets\cite{art:9000215}. 
For these Hamiltonians we consider a partition of two half spaces as shown in figure \ref{fig:front}.  Region A consists of a half space containing one atom, and region B consists of the half space containing the other atom. 
 The  Renyi entropy of several dimers at their equilibrium geometries is shown in figure \ref{fig:molecules} using the three methods. In figure \ref{fig:c-ent} we plot the Renyi entropy as a function of atomic separation for the C$_2$ dimer over a range in which the covalent bonds between the atoms are stretched and broken.

\begin{figure}[htp]
\centering
\includegraphics[scale=0.4]{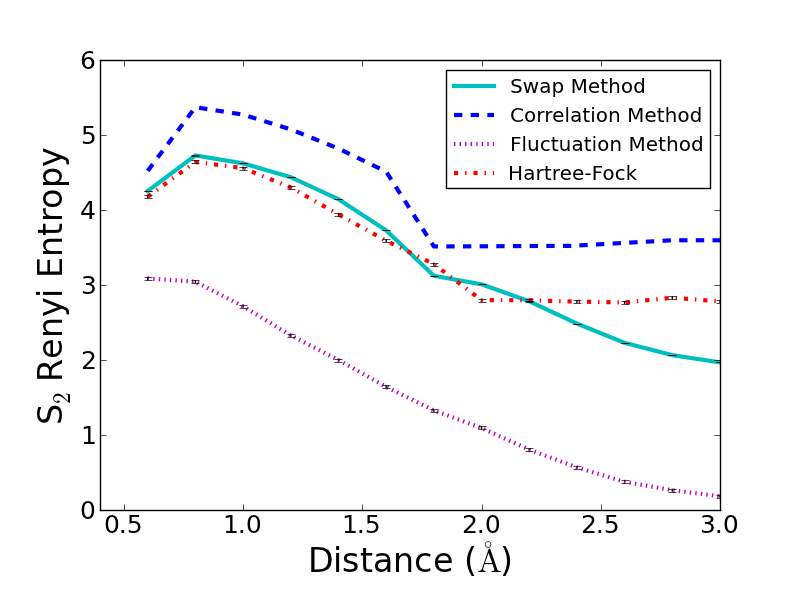}
\caption{S$_2$ Renyi entropy calculated for the C$_{2}$ dimer for the multi-determinant wave functions. The Hartree-Fock values are plotted as well for reference. The equilibrium distance is at 1.2 $\AA$. \label{fig:c-ent}}
\end{figure}


For these systems, Hartree-Fock wave functions are not always accurate representations of the ground state wave function. 
 We have tested a variety of different wave functions and determined that multi-determinant wave functions using a range of 100-1000 determinants, depending on the system, to be quite accurate\cite{gamess-1,art:9000344,art:9000215}. 
These wave functions are sufficiently compact to include all determinants in the VMC calculation, and in this form it is straightforward to compute all three techniques for calculating the Renyi entropies.

The results in figures \ref{fig:molecules} and \ref{fig:c-ent} show that the correlation and fluctuation methods capture some of the qualitative behavior of the Renyi entropy but not in the limit of large atomic separation of C$_{2}$.  It is interesting that these two methods consistently underestimates or overestimates the Renyi entropy for all the systems considered here.  Neither  estimate is consistently accurate; however when considered together they appear to bracket the actual value of the Renyi entropy.  This can be partly understood from known properties of the fluctuation method. It has been shown that estimates of S$_{1}$, from the fluctuation method are strictly less than or equal to the actual value of S$_{1}$\cite{measure-4}.  We observe this bracketing holds for all Renyi entropy calculations considered in this work, and it would certainly be of interest to explore its generality further.   

In the region of the equilibrium geometry for C$_{2}$ the Renyi entropy of the multi-determinant wave function is in relatively good agreement with the Renyi entropy of the Hartree-Fock wave function.  These results show the Renyi entropy increases as the atoms are squeezed past the equilibrium geometry followed by a maximum and reduction in the entropy.  The agreement between the multi-determinant and the Hartree-Fock wave functions breaks down as the separation between the atoms increases, which is also the region in which the fluctuation and correlation methods show large deviations from the benchmark swap results.  The Renyi entropy from the fluctuation method decays quickly when compared to the swap operator.  In contrast, the correlation method does not allow the electrons to disentangle in the bond breaking region.  In this limit it is expected that the fluctuation method significantly underestimates the Renyi entropy as electron number fluctuations between subsystems are only part of the Renyi entropy for an interacting system. Two regions can have finite entanglement without fermions exchanging between them such as in a dispersion bonded system.

Interpreting the value of the Renyi entropy can lead to insight into properties of molecular systems and here we consider how it might be related to the bonding properties.  We consider the cases of C$_{2}$ and N$_{2}$ at their equilibrium geometries, where the Renyi entropy is slightly greater than 4.0 for both molecules, and we ask what is the relevance of the magnitude of these Renyi entropies.  We develop an interpretation of these values by considering the entanglement of a single non-interacting electron divided evenly in half by region A and B for example the case of a particle in a box that is split by the midpoint of the box.  The results of this model is a density matrix $\rho_{A}$ that has two eigenvalues (1/2,1/2) which yields a value ln(2) for the S$_{2}$ Renyi entropy.  For our analysis it is important to determine the entanglement Hamiltonian, which in this case has a single eigenvalue of zero.     
Extending this analysis to other single particle Hamiltonians, it can be shown for all reasonable single particle wave functions that the entanglement Hamiltonian will have an eigenvalue of zero if the chosen regions divide the probability density equally. 

If we model a bond with any single particle model in which probability density is split in half, the signature result is a zero eigenvalue of the entanglement Hamiltonian.  If one takes the associated Renyi entropy value, ln(2), as a normalization of the number of electrons being covalently shared between two subsystems, then we can estimate the number of covalent bonds in the system with N$_{bond}$ = S$_{2}$/(2$\cdot$ln(2)).  Conventionally a bond is occupied by two electrons, which is why a factor of two is included in this definition.  For C$_{2}$ and N$_{2}$ at their respective equilibrium geometries this is suggestive of three covalent bonds.  This is not the standard two bonds that one would expect from traditional bond order theory of C$_{2}$ \cite{bo-1,bo-2,bo-3}. We note that the bond order of C$_{2}$ is still an open question and has been the subject of recent studies\cite{cbond-1,cbond-2}. As for the other dimers, the results for Li$_{2}$ and F$_{2}$ are in agreement with conventional bond order values. However Be$_{2}$ is well known to have unusual bonding characteristics, and its conventional bond order is zero\cite{bbond-1}. This is problematic as Be$_{2}$ is known to be a stable molecule and our result of a partial bond for Be$_{2}$ is what should be expected from physical considerations.

  It is important to further extend this analysis as the Renyi entropy only provides an average over some of the information in the density matrix $\rho_{A}$.  In particular not all contributions to the Renyi entropy are necessarily due to perfectly covalent bonds.  More insight can be gained by considering the eigenvalues of the entanglement Hamiltonian\cite{op-2}.  For the single determinant Hartree-Fock wave functions the eigenvalue spectrum can be calculated directly with the correlation method.  In this case the $2^{N}$ eigenvalues of $\rho_{A}$ are mapped on to the $N$ eigenvalues of the entanglement Hamiltonian.

\begin{figure}[tbp]
\centering
\includegraphics[scale=0.4]{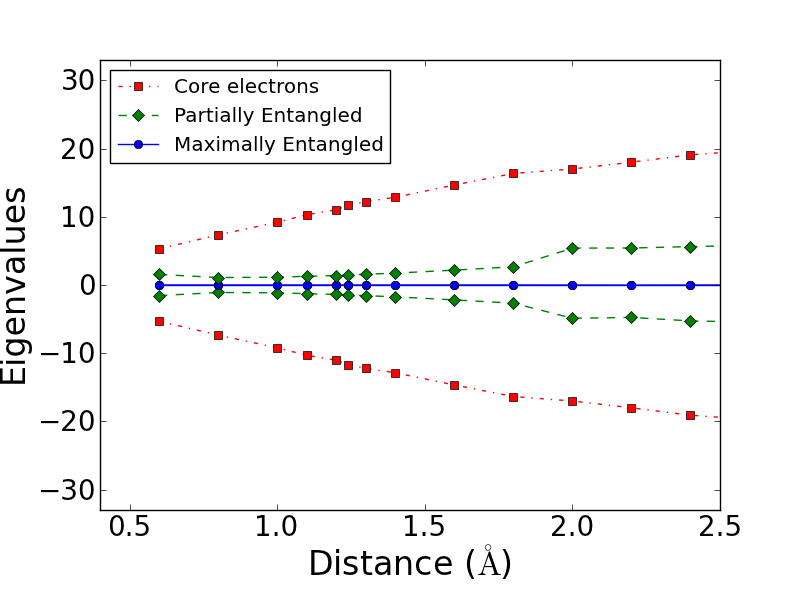}
\caption{Eigenvalues of the Hartree-Fock entanglement Hamiltonian calculated for the C$_{2}$ dimer.  There are twelve eigenvalues total, all non-zero eigenvalues are doubly degenerate and the zero eigenvalue is quadruply degenerate.  Eigenvalues with large magnitude contribute less to the entanglement and the four eigenvalues with the largest magnitudes do not contribute to the entanglement significantly at any point along the binding curve.  \label{fig:c-spec}}
\end{figure}

 We proceed by observing that in the bonding region of C$_{2}$ there is good agreement between the Hartree-Fock and multi-determinant Renyi entropies, as seen in figure \ref{fig:c-ent}. This indicates that the spectrum produced by the Hartree-Fock entanglement Hamiltonian may give an accurate representation of the multi-determinant wave function spectrum.   The Hartree-Fock entanglement Hamiltonian for  C$_{2}$ and N$_{2}$ contain several zero eigenvalues.  For C$_{2}$ there are 4 zero eigenvalues, and for N$_{2}$ there are 6 zero eigenvalues, which is what one would expect in traditional measurements of bond order for these molecules. The eigenvalue spectrum for C$_{2}$ is plotted in figure \ref{fig:c-spec}.  These two systems differ however in terms of the remaining eigenvalues.  In the case of N$_{2}$ the rest of the eigenvalues are large and do not contribute significantly to the entanglement, whereas in C$_{2}$ there are four more non-trivial values in the eigenvalue spectrum as shown in figure \ref{fig:c-spec} by the two curves with diamond symbols.  

Our results suggest that there are two covalent bonds for C$_{2}$ and three covalent bonds for N$_{2}$.  The other valence electrons in C$_{2}$ appear to be strongly covalent as they make significant contributions to the Renyi entropy. It is likely these eigenvalues are a signature of the two extra bonds recently predicted in a study on bonding in C$_{2}$\cite{cbond-1}.  In this study an energy was determined for each of the four bonds in C$_{2}$ resulting in three energetically strong bonds and one weak bond.  This is in contrast to our entanglement results which show two perfectly covalent bonds and two partially covalent bonds.  Differences between these results likely arise from our use of the entanglement Hamiltonian for the Hartree-Fock wave function.  It is possible to determine the eigenvalues of $\rho_{A}$ for multi-determinant wave functions through the calculation of many higher order Renyi entropies, S$_{n}$\cite{measure-1}. However, this requires significant computational resources for C$_{2}$ and developing more efficient techniques for this purpose is certainly an interesting direction for future research.

In conclusion we report one of the first \textit{ab initio} calculations of Renyi entropy including Coulomb interactions while demonstrating the effect of interactions on proposed experimental measurements of entanglement.  In general the fluctuation and correlation methods do not appear to accurately determine the Renyi entropy.  However for the systems under consideration here these methods bracket the Renyi entropy.

Intuitively the use of entanglement appears to be a natural way to characterize bonds and our results suggest there are many possibilities for applying these techniques to chemical systems.  We have shown that entanglement can be used to help describe covalent bonding in molecules and may be useful to describe other bonding situations.  Other molecular systems of interest would be ionic and van der Waals bonded systems as well as transition metal dimers where the bonding properties can be difficult to characterize.

 In the case of more complicated molecules it will be important to consider the use of entanglement properties to generate partitions in space, similar to what is done with Bader analysis\cite{bader:book}.  Finally we note that that there is large interest in using density functional theory for calculating entanglement entropy, where methods are already being developed\cite{dftent-1}.  It is clear that the results presented in this work as well as those from future QMC studies using the techniques described here can serve as benchmarks for such developments. 

\section{ACKNOWLEDGMENTS}
We would like to acknowledge useful discussions with David Ceperley, Sarang Gopalakrishnan, Jonathan Moussa, Noah Bray-Ali and Ann Kallin. This work was supported by the National Science Foundation under grant OCI-0904572.  

%

\section{Supplemental material}

We have used several different techniques to calculate the entanglement properties in this work.  
Here we describe our implementation and usage of these algorithms.

\subsection{Particle Fluctuation Method}
  
The fluctuation method is calculated with moments (and the related cumulants), of the particle number operator, which are straightforward to estimate in a VMC calculation.  In non-interacting systems the relationship between the cumulants and the S$_{2}$ Renyi entropy is given as follows
\begin{equation}
S_{2}= \lim_{K \rightarrow \infty} \sum_{n=1}^{K+1}\alpha_{n}(K)C_{n}
\end{equation}
with the cumulants, $C_{n}$, and a set of constants, $\alpha_{n}(K)$, that depend on the cutoff used in the expansion, K.  Similar expressions also exist for the rest of the Renyi entropies, and expressions for the coefficients $\alpha_{n}(K)$ have been determined\cite{measure-1}.  The S$_{2}$ Renyi entropy converges quickly with the number of terms kept in our results.  We calculate the first six cumulant terms for use in this expansion in our VMC calculations. For testing the cumulant cutoff we compared against the swap operator for the Hartree-Fock calculations discussed in this work. The number of terms we included is more than sufficient to attain agreement between the two results within the error bars of their estimates.


\subsection{Correlation Method}  
Calculating correlation functions of a quadratic Hamiltonian is another approach to calculating entanglement entropy that has been used significantly for calculations on lattice models.  These calculations are performed by mapping the Hamiltonian of the full system to the entanglement Hamiltonian, which is defined as a quadratic Hamiltonian that produces the density matrix of the subsystem.
\begin{equation}
\label{eqn:entrho}
\rho_{A} = \kappa e^{-\mathcal{H}_{ent}}.
\end{equation}
The values $\kappa$ is a normalization that we use to fix $Tr(\rho_{A})=1$.
The correlation matrix is defined as
\begin{equation}
C_{ij} = \langle c_{i}^{\dag}c_{j} \rangle = Tr(\gamma_{1}c_{i}^{\dag}c_{j}).
\end{equation}
The entanglement Hamiltonian is then given by the correlation matrix and the identity matrix $\mathbf{I} $\cite{redmat-1}, 
\begin{equation}
H_{ent} = ln[\frac{\mathbf{I}-\mathbf{C}}{\mathbf{C}}].
\end{equation}
The eigenvalues of the entanglement Hamiltonian can be used to calculate the eigenvalues of $\rho_{A}$ by equation \ref{eqn:entrho}.

We were able to efficiently apply these methods to Hartree-Fock wave functions, for which this method is exact, and for multi-determinant wave functions, in which the correlation matrices only approximately determine the entanglement properties.  The correlation matrices of interest for these wave functions can be calculated from the natural orbitals by projection onto a real space grid that exists only in the subregion of interest.  The straightforward way of doing this involves diagonalizing $C_{ij}$ on an $l$x$l$ matrix, where $l$ is the number of grid points in region A.  The matrix $C_{ij}$ has a rank much smaller than the number of grid points, and a method
that scales significantly better with the size of the grid can be performed by generating the following matrix from the natural orbitals $u_{i}(r_{j})$ and a grid  $r_{j}$ in region A,
\begin{equation}
M_{ij}=f_{i}^{1/2}u_{i}(r_{j}).
\end{equation}
The correlation matrix is formed by matrix multiplication $\mathbf{M}\mathbf{M}^{T}$.  Instead of diagonalizing this matrix, one can calculate the singular value decomposition of $M_{ij}$. By squaring the resulting singular values one recovers the eigenvalues of the correlation matrix.  This has the advantage of reducing the computational scaling with the number of grid points.  We used a fine cubic grid of a minimum of 1 million points to minimize errors, and the sum of the eigenvalues of $\rho_{A}$ was normalized to one.

The $f_{i}$ in the case of Hartree-Fock are equal to 1 for occupied orbitals and 0 for unoccupied orbitals.  More generally they are equal to the eigenvalues of the natural orbitals for a given wave function.  This technique can be used with any wave function based method, in which the natural orbitals can be calculated.  The natural orbitals can also be calculated within QMC, although this was not our approach in this work.  

 It has been shown previously that only a single sum over the eigenstates $\epsilon_{i}$ of the entanglement Hamiltonian is needed to calculate the entanglement entropy \cite{redmat-1}. A similar iterative formula over all the eigenstates can also be used to calculate the Renyi entropy $S_{n}$. This is done by calculating a set of dummy variables $w_{n}$ and $\kappa_{n}$. We start by setting $w_{0} = 0$ and $\kappa_{0} = 1$ and we calculate $w_{i+1}$ and $\kappa_{i+1}$ iteratively, 
\begin{eqnarray}
w_{i+1} = e^{-n\epsilon_{i}}(1+w_{i})+w_{i}\nonumber\\
\kappa_{i+1} = \kappa_{i}(1+e^{-\epsilon_{i}}). \label{eqn:ren1}
\end{eqnarray}
After calculating $w_{N_{tot}}$ and $\kappa_{N_{tot}}$, the Renyi entropy is given by
\begin{equation}
S_{n} = \frac{w_{N_{tot}}+1}{(\kappa_{N_{tot}})^{n}}.
\label{eqn:ren2}
\end{equation}
The value $n$ is the order of the Renyi entropy calculated and N$_{tot}$ is the number of eigenvalues. The number of non-trivial eigenvalues will at most be equal to the number of electrons in the case of the Hartree-Fock wave functions, or in the more general case it will be no larger than the number of natural orbitals used to calculate the correlation matrix. 

  This brings up an ambiguity of applying the correlation method to a wave function that is not a single determinant.  In principle the ground state correlation functions can be measured experimentally, and will agree with the correlation functions we calculate for high quality wave functions.  Using such a correlation function yields a number of eigenvalues for the correlation matrix that exceeds the number of electrons in the initial system.  In this work we applied equations \ref{eqn:ren1} and \ref{eqn:ren2} to the full set of non-trivial eigenvalues that come from such a correlation matrix, but this is not the only approach for calculating Renyi entropies when dealing with a correlation matrix for an interacting system.

\subsection{Swap operator method in quantum Monte Carlo}
Finally we extended a recent QMC method to calculate Renyi entropies exactly from a given wave function, to work on our molecular \textit{ab initio} Hamiltonians.  
A recent series of papers have shown that using a swap operator within lattice QMC calculations yields a technique in which Renyi entropies can be calculated\cite{renyi-1,renyi-2,renyi-3,renyi-4}.  The operation of the swap operator is defined on a wave function that is written in terms of complete basis sets of regions A and B, $|\alpha \rangle$ and $|\beta \rangle$,  as follows $|\Psi\rangle = \sum C_{\alpha,\beta}|\alpha\rangle |\beta\rangle$.  
The swap operator is defined in a space of the wave function that is a tensor product with itself as follows 
\begin{eqnarray}
swap_{A}(\sum_{\alpha_{1},\beta_{1}}C_{\alpha_{1},\beta_{1}}|\alpha_{1}\rangle|\beta_{1}\rangle) \otimes 
(\sum_{\alpha_{2},\beta_{2}}D_{\alpha_{2},\beta_{2}}|\alpha_{2}\rangle|\beta_{2}\rangle) \nonumber \\
= \sum_{\alpha_{1},\beta_{1}}C_{\alpha_{1},\beta_{1}} 
\sum_{\alpha_{2},\beta_{2}}D_{\alpha_{2},\beta_{12}}(|\alpha_{2}\rangle|\beta_{1}\rangle 
\otimes |\alpha_{1}\rangle|\beta_{2}\rangle ).\nonumber \\
\end{eqnarray}

The expectation value of the swap operator is related to $\rho_{A}$ by the following,
\begin{equation}
 Tr((\rho_{A})^{2}) = \frac{\langle \Psi_{T} \otimes \Psi_{T}| swap_{A} |\Psi_{T} \otimes \Psi_{T} \rangle}{\langle \Psi_{T} \otimes \Psi_{T}|\Psi_{T} \otimes \Psi_{T} \rangle}.
\end{equation}
  This operator can be sampled with pairs of walkers independently sampled from $|\Psi_{T}|^{2}$ and the results can be used to calculate the Renyi entropy, S$_{2}$ = $-ln(Tr(\rho_{a})^{2})$. 
In terms of the walker coordinates the expectation value of the swap operator is  given by equation \ref{eqn:estimator}.
\begin{figure*}
\begin{eqnarray}
\label{eqn:estimator}
 \langle \Psi_{T} \otimes \Psi_{T}| swap_{A} |\Psi_{T} \otimes \Psi_{T} \rangle 
= \int d\mathbf{x}_{1}d\mathbf{x}_{2} \cdots d\mathbf{x}_{2N} \Psi^{*}_{T}(R(\alpha_{1},\beta_{1}))\Psi^{*}_{T}(R(\alpha_{2},\beta_{2}))\Psi_{T}(R(\alpha_{2},\beta_{1}))\Psi_{T}(R(\alpha_{1},\beta_{2})) \nonumber \\
= \int d\mathbf{x}_{1}d\mathbf{x}_{2} \cdots d\mathbf{x}_{2N} |\Psi_{T}(R(\alpha_{1},\beta_{1}))|^{2}|\Psi_{T}(R(\alpha_{2},\beta_{2}))|^{2}\frac{\Psi_{T}(R(\alpha_{2},\beta_{1}))\Psi_{T}(R(\alpha_{1},\beta_{2}))}
{\Psi_{T}(R(\alpha_{1},\beta_{1}))\Psi_{T}(R(\alpha_{2},\beta_{2}))}
\end{eqnarray}
\end{figure*}
From these definitions the VMC operator can be identified as 
\begin{equation}
O(\alpha_{1},\alpha_{2},\beta_{1},\beta_{2}) = \frac{\Psi_{T}(R(\alpha_{2},\beta_{1}))\Psi_{T}(R(\alpha_{1},\beta_{2}))}
{\Psi_{T}(R(\alpha_{1},\beta_{1}))\Psi_{T}(R(\alpha_{2},\beta_{2}))}
\end{equation}
In these equations we have used the notation $R(\alpha,\beta)$, which are the coordinates of a walker.  The swap operator is performed over two walker configurations, and thus the subscripts are used to identify whether the coordinates are from the first or second walker.  All the coordinates in region A are associated with $\alpha$ and all the coordinates in region B are associated with $\beta$.  For our wave functions in which spin and particle number is held fixed, if the swap operator causes a change in spin or particle number, then the expectation values of the swap operator for that pair of 
configurations is zero.

\end{document}